\documentclass{article}

\renewcommand{\abstract}[1]{{ \footnotesize \noindent {\bf
Abstract} #1 \\}}
\renewcommand{\author}[1]{\subsubsection*{#1}}
\newcommand{\address}[1]{\subsubsection*{\it#1}}

\usepackage{epsf1990}

\newcommand\gtapprox[0]{\,\lower.6ex\hbox{$\buildrel >\over \sim$} \, }
\newcommand\ltapprox[0]{\,\lower.6ex\hbox{$\buildrel <\over \sim$} \, }

\newcommand\hkpc[0]{\mbox{$\,h^{-1}\,$kpc}}
\newcommand\hGpc[0]{\mbox{$h^{-1}\,$Gpc}}

\newcommand\Omm[0]{\Omega_{\mbox{\rm \small m}}}
\newcommand\rinj[0]{{r}_{\mbox{\rm \small inj}}} 

\newcommand\cqg[0]{Class. Quant. Grav.}         
\newcommand\mnras[0]{Month. Not. R. Astron. Soc.}

\newcommand\joref[5]{#1, #5, {#2, }{#3, } #4}
\newcommand\epref[3]{#1, #3, {#2}}

\begin{document}

\title*{Observational Constraints on the Topology (Global
Geometry) of the Universe}

\author{Boudewijn F. Roukema}

\address{LUTH -- Observatoire de Paris--Meudon, 5, place Jules Janssen, 
F-92.195, Meudon Cedex, France (Boud.Roukema@obspm.fr)\\ 
Nicolaus Copernicus Astronomical Center, 
ul. Bartycka 18, P-00-716 Warsaw, Poland\\
University of Warsaw, Krakowskie Przedmie\'scie 26/28, 
P-00-927 Warsaw, Poland}

\abstract{The Universe is a physical object.  Physical objects
have shapes and sizes.  General relativity is
insufficient to describe the global shape and size of
the Universe: the Hilbert-Einstein equations only treat
limiting quantities towards an arbitrary point.
Empirical work on measuring the shape and size of the
Universe (formally: the ``3-manifold of the spatial
hypersurface at constant cosmological time'', and,
e.g. the ``injectivity diameter'' respectively) has
progressed significantly in the late 1980's and the
1990's, using observational catalogues of galaxy
clusters, of quasars and of the microwave background, 
though the analyses are still hindered by simplifying 
(and often observationally unsupported) assumptions.
A review of the different observational strategies 
and claimed constraints was presented at the meeting.}


\section{Introduction}
The Universe is a physical object.  Physical objects have shapes and
sizes.

So, a major goal of observational cosmology is to measure the shape
and size of the Universe (within appropriate mathematical theory
relating to geometry), or else to convincingly show that these are
unmeasurable.  The alternative, to suppose that the Universe is a
spiritual object, without a shape or size, is not part of the domain
of science.

\subsection{Relativity and geometry}
General relativity relates local geometry to the physical content of
the Universe. However, it is insufficient to describe the global shape
and size of the Universe: the Hilbert-Einstein equations only treat
limiting quantities towards an arbitrary point, i.e. they are local.

An extension of general relativity, for example, a theory of 
quantum gravity, should relate the physical content of the Universe
to its global shape and size. J\"urgen Ehlers [private communication]
pointed out that in this sense, one could say that general relativity,
as a theory relating gravity to geometry, is incomplete, so that 
while the global geometry of the Universe is independent of the present
theory of general relativity, it could be said to be constrained by
a more complete form of general relativity, in a theory yet to be found 
and/or agreed upon.

\subsection{Observational detection: topological lensing}
Just as general relativity and local perturbations in geometry lead to
the observable phenomenon of {\em gravitational} lensing, whatever theory
extends general relativity, yielding constraints on global geometry,
would reveal itself by {\em topological} lensing \cite{Wambs02}.

This is the basic principle common to (nearly) all suggested
techniques of detecting global geometry. (See 2.B.i and 2.B.ii of
\cite{Rouk02} for the only techniques known to this author which are
independent of this principle.)
Just as gravitational lensing
is caused by multiple geodesics from an object to the observer,
topological lensing would also be caused by multiple geodesics to
the observer. In the former case, the geodesics only differ very
slightly, over a small portion of their length --- due to the
gravitational distortion induced by an intervening massive object. But
in the latter case, the geodesics are, in general, of very different lengths 
and point in very different directions, since they are simply two different 
ways of crossing the Universe between two points.

The ``object'' observed may either be a gravitationally collapsed, luminous 
astrophysical object (methods A.i listed in \cite{Rouk02}), 
or a ``patch'' of photon emitting plasma seen as a
fluctuation in the cosmic microwave background (methods A.ii listed in 
\cite{Rouk02}).
Due to the pioneering stage of this research, terminology and approaches 
to classifying the different methods of applying of this principle still vary
somewhat.

\newcommand\fgaia{\begin{figure*} 
\centerline{\epsfxsize=\textwidth
{\epsfbox[-104  105  596 586]{"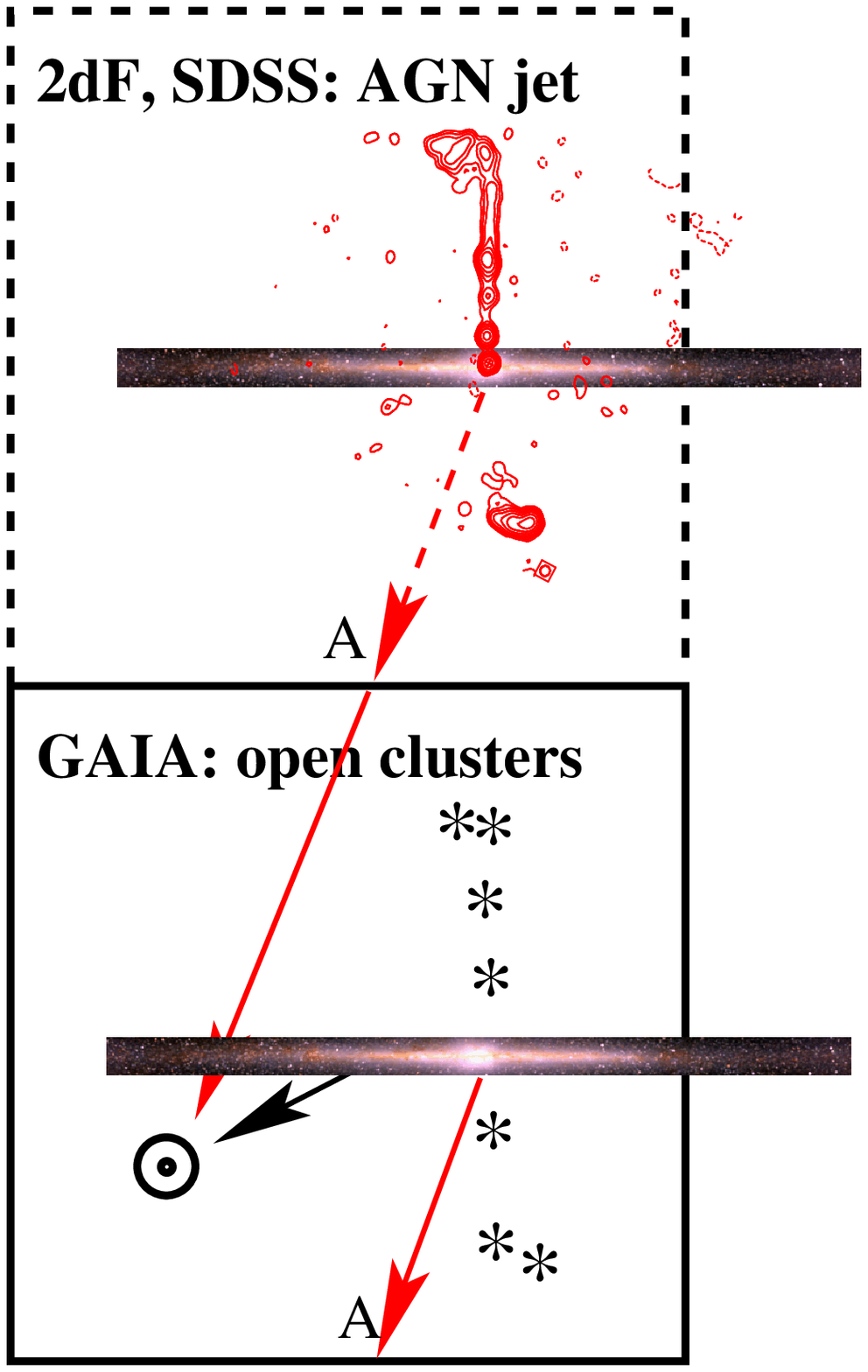"} }
}
\caption{\label{f-gaia} 
For explanatory purposes, an exaggeratedly small toroidal universe,
about 16kpc in size, is shown. As explained in \protect\cite{Rouk00d}, 
a two-dimensional, flat global universe can be thought of either
as (i), a 2-torus placed in ordinary Euclidean 3-space, but then given
an intrinsically flat metric, (ii) a ``cut-open'' 2-torus, i.e. a 
rectangle of which opposite sides are identified with one another, 
(called the ``fundamental polyhedron'') or (iii) a tiling of the Euclidean
2-plane by multiple copies of the rectangle (called the ``universal 
covering space'').
The solid outline (lower square) includes the entire physical
universe (fundamental polyhedron). Method (ii) of thinking of 
the space can be applied by ignoring everything outside this solid outline.
The dark arrow shows the geodesic from the Galactic Centre
to the observer at the Sun.
The gray arrow shows the (long time delay) loop around the universe, where
the light from the Galactic Centre takes a lot longer to arrive at the
observer. Because the time delay is much bigger, it may correspond to 
the delay calculated from GAIA data for a ``high'' redshift (early epoch), 
extragalactic, AGN
phase image of the Galaxy.
This is illustrated by extending to the 
dotted outline (upper square), 
showing a topological 
image of the universe, in apparent space, i.e thinking of the Universe
as a tiling (iii). The AGN phase is schematically shown by 
a double lobe radio jet $\sim10$kpc in full length. 
Returning to the lower half of the figure, 
open star clusters are shown by asterisks following the shape of
the ``high'' redshift jet. These represent star clusters formed during the
AGN phase, which would be observable ``today'' by GAIA. 
In reality, they would no longer occupy the ``vertical'' axis of the Galaxy; 
they would probably 
have been through several orbits since the AGN phase during
which they formed.
}
\end{figure*} }

\newcommand\frlagn{\begin{figure*} 
\centerline{\epsfxsize=\textwidth
{\epsfbox[20 125 635 747]{"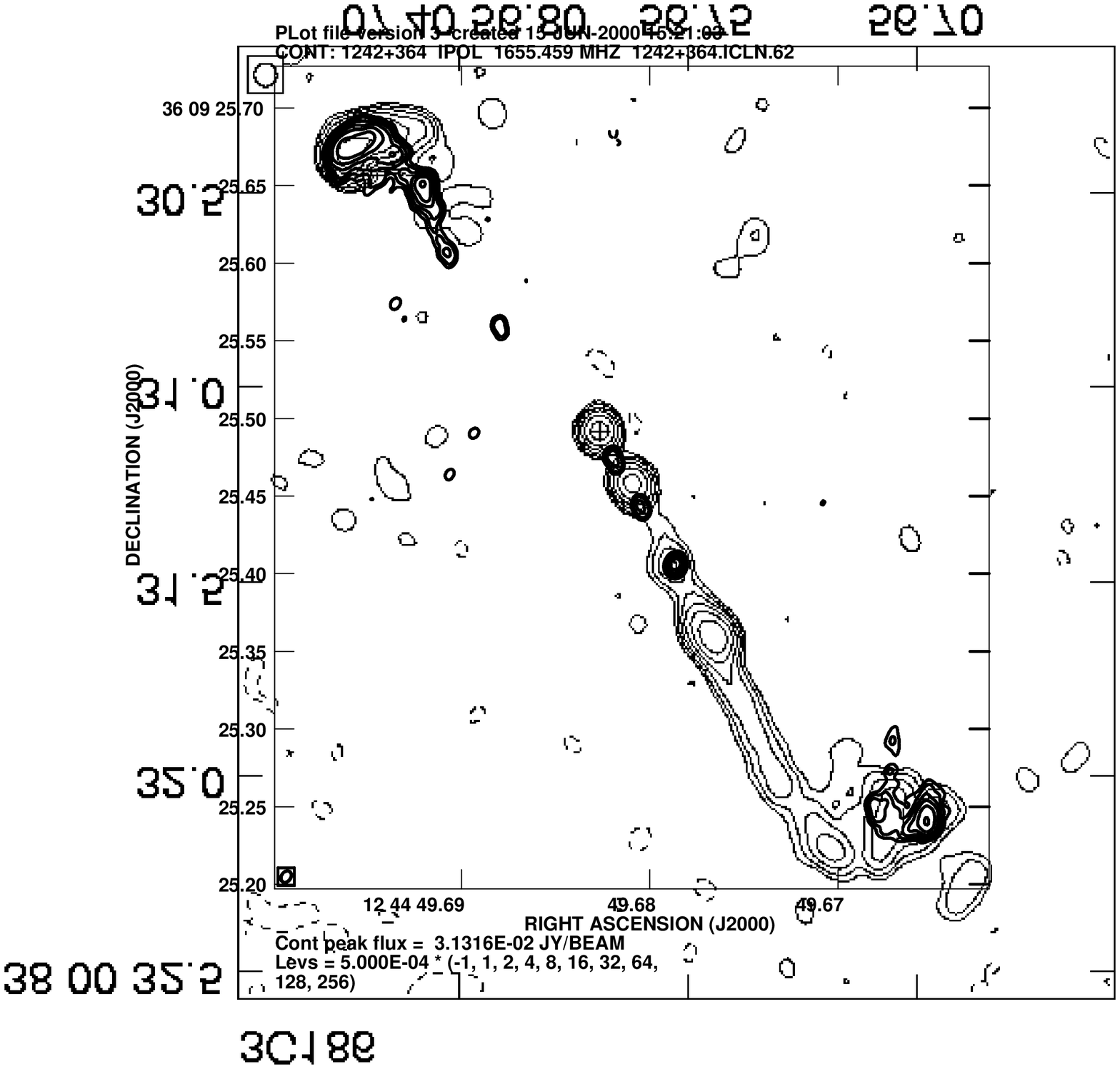"} }
}
\caption{\label{f-rlagn} 
The two strikingly similar radio-loud AGNs. The image 
of 4C+36.21 (at 18cm) is shown in heavy contours; the image of 3C186
(at 6cm) is shown in light contours and reflected North-South. 
Either (i) these two images are of physically distinct objects
which just happen to show a 
similar physical process or (ii) they are two topologically
lensed images of a single object, separated from one another in 
comoving space by a ``loop around'' the Universe.
}
\end{figure*} }

\section[Recommended reading]{A non-exhaustive list of 
recommended reading on background mathematics (geometry/topology), 
(lack of) physical 
theory and observational strategies}

Apart from articles based on workshops in \cite{Stark98} and \cite{BR99}, 
the following may help guide the reader through the rapidly expanding
literature.

\subsection{Geometry/topology}
For the background mathematics (and some comments on observational
strategies), \cite{LaLu95} is recommended, but has been complemented
recently by a thorough article on multiply connected spherical spaces
\cite{Gaus02}, which have become relevant due to increasing evidence
that the observable Universe is approximately flat, just as the
surface of the Earth is nearly flat.  The curvature radius $R_C$ of a
section of the Earth's surface of radius $R_H \sim
6400$~km or smaller satisfies $R_C \gtapprox R_H$; similarly, the
curvature radius $R_C$ of the observable Universe is estimated in
comoving units as $R_C \gtapprox R_H \approx 10$~{\hGpc}, where $R_H$
is the horizon radius of the Universe and the {\em local} cosmological
parameters, $\Omm$ and $\Omega_\Lambda$, the density parameter and the
cosmological constant respectively, are $(\Omm=0.3, \Omega_\Lambda=0.7)$.

\subsection{(Lack of) physical theory}
For a shorter description of the background mathematics, but also 
some references to the beginnings of theoretical work which 
could be useful for a physical theory of global geometry, \cite{LR99} 
is recommended. (This also includes a short historical introduction.)
Ideas on topological evolution during the quantum epoch \cite{DowS98} 
and a diverse range of physical approaches in {\S}VI of \cite{Fag02} 
are just a few examples of theoretical work.

\subsection{Observational strategies}
Definitions of the more formal terms corresponding to ``shape'' and ``size'',
e.g. the ``3-manifold of the spatial
hypersurface at constant cosmological time'' for ``shape'', and
the ``injectivity diameter'' (twice the injectivity radius) for ``size'', 
are illustrated in fig.~10, \S5.1 of \cite{LR99}, using the terminology
of \cite{Corn98a}.

Most of the empirical work has (understandably) been done
independently of any physical theory of global geometry, i.e. has just
assumed a standard Friedmann-Lema\^{\i}tre-Robertson-Walker metric. 
For a rapid introduction to this approach, newcomers to the field
might want to skip straight to \S5.1 of \cite{LR99}, and fill in later
on the fuller mathematical and historical background.

For an overview and brief description of how the principle of multiple
imaging is applied in practice, (i) to collapsed objects spread through
three-dimensional comoving space
and (ii) to the cosmic microwave background, which would uniquely be in 
two-dimensional comoving space if the cosmological constant were zero and
the density parameter unity, see \cite{Rouk00d}. 
Although several 
methodical developments and observational analyses have been carried 
out by various groups since \cite{LR99}, the observational constraints 
on the size of the Universe, i.e. on the injectivity diameter, remain
essentially unchanged from the scales listed for the different approaches
in table~2 of \S5.2 of \cite{LR99}, i.e. $2\rinj \gtapprox 1$~{\hGpc}.

The history of the search for exoplanets in the 1990's suggests
that the inclusion of a ``reasonable'' but uncertain theoretical assumption
(that a planet massive enough to be detectable could not occur close to
its parent star, with an orbital period of only a few days) 
may lead to ignorance of an astrophysical 
discovery present 
in existing observational data. Nevertheless, as a strategical 
choice it is valid, as long as strong claims are not made.

This is the case in most of the cosmic microwave background analyses
for cosmic topology, listed as A.ii.3 in \cite{Rouk02}. 

A brief discussion of what the assumptions are and why they limit 
the generality of the conclusions of those analyses is provided in
\S1.2 of \cite{Rouk00a}.

A new major review on cosmic topology, mostly focussing on cosmic 
microwave background methods, is \cite{Levin01}.

For the more observationally minded, two applications of approach 
A.i.3, 
using physical characteristics of individual objects in order
to detect candidate topologically lensed images,
published more recently than refs
\cite{Rouk00d} and \cite{Rouk02}, are presented here
as illustrations of some of the more direct observational 
approaches possible.

\fgaia

\section[GAIA: the Milky Way as an AGN]{GAIA: 
the Milky Way as the ultimate extragalactic source}

By the end of the decade, it is planned to launch a satellite, GAIA, which
will make parallax measurements of a billion stars in the Milky
Way. This should lead to unprecedented
understanding of how the Milky Way formed, and should make it possible
to estimate dates of important past events in the history of the Galaxy,
including merger events with small neighbouring galaxies, and major 
events of ``nuclear'' activity, i.e. of times when our galaxy would have
appeared as a quasar or other Active Galactic Nucleus galaxy (AGN) if seen 
from afar. (In fact, evidence is mounting that 
mergers and AGN phases probably coincide.)

Since AGNs are typically seen to high redshifts, e.g. $z\sim2$, the precise
dating of past AGN events of the Galaxy would indicate narrow bands of 
redshift in which topologically lensed images of our Galaxy would be
seen. This could lead either to detection of candidate generators (a 
generator is a path 
joining topologically lensed images, corresponding to a single 
``loop around'' the Universe), or to increased confidence in lower 
limits to $2\rinj$ (apart from caveats due to AGNs being missed in the
plane of the Galaxy).

Fig.~\ref{f-gaia} illustrates the observational situation, and also 
shows an example of some of the basic elements of global geometry in 
a Friedmann-Lema\^{\i}tre-Robertson-Walker Universe. For developing one's 
basic intuition of observational cosmic topology, it is recommended to 
be able to switch between thinking in modes (i), (ii) and (iii) 
as listed in the figure and in, e.g. \cite{Rouk00d}. Mode (i) is probably
the most intuitive for the 2-dimensional case, but difficult for 3-dimensional
space; mode (ii) is probably easiest for thinking of the physics; and
mode (iii) is generally best for analysing observations.

So GAIA might just possibly 
turn out to be a powerful probe for observational cosmology. 
See \cite{Wich00} for a recent discussion of searching for high redshift
images of the Galaxy, and \cite{Rouk01a} for brief comments on the 
relevance of GAIA.

\frlagn

\section[Similar RLAGN morphology]{Similar 
morphologies of radio-loud active galactic nuclei (AGNs)}

Although topologically lensed pairs of images would in general
occur at very different redshifts, there are certain cases where
the redshifts can be very nearly equal. The ``matched circles principle'' 
(\cite{Corn96,Corn98b}; see fig.~2 of \cite{Rouk00d} for an explanation
of the principle and fig.~3 of \cite{Rouk00d} for examples of matching
and non-matching circles in 4-year COBE data), adapted in the obvious
way for an arbitrary sub-horizon redshift, defines these cases.

A striking coincidence in the morphologies of two radio-loud, 
double-lobed, compact steep spectrum AGNs, 3C186 and 4C+36.21, 
yields a good illustration of the falsifiability of specific candidate
generators of the global geometry of the Universe \cite{RMBS01}.

The redshift of 3C186 is known: $z=1.063$,  
but the redshift of 4C+36.21 is unknown. 
These two images could only be
two topological images of a single RLAGN if the redshift of 
4C+36.21 were to lie in the very small interval which gives a physically 
reasonable expansion speed 
for the jet (positive and not slower than about $0.01c$).

4C+36.21 is seen with a linear size in
proper units of about 1.6{\hkpc}, and 3C186 is 6.5{\hkpc} in size,
so that the 4C+36.21 image must be just slightly
earlier in time than 3C186, but not too much earlier, if the 
identity hypothesis were to be correct.

The measured redshift of 4C+36.21 would have to lie in the
very narrow range $1.0630< z \ltapprox 1.0635$. 

A spectroscopic estimate of the redshift of 4C+36.21 is planned.
A redshift outside of this range would clearly refute the hypothesis.

A redshift within the required range would be exciting, but 
would require many more observational tests before it could be
considered to provide a serious candidate estimate (roughly 
1~{\hGpc}) of the size of the Universe.

\section{Conclusion}

Although the subject of cosmic topology is just over a century 
old (\cite{Schw00}; it predates general relativity), it was only 
in the late 1980's and the 1990's that significant 
observational attention started being paid to the subject. With 
estimates of the local cosmological parameters finally converging on
consistent values, the natural followup is 
the quest for measuring global cosmological parameters.


\begin{thebibliography}{99}
%
%
%


\bibitem{BR99} 
Blanl{\oe}il V., Roukema B.~F., 2000, 
E-Proceedings of the 
{\it Cosmological Topology in Paris 1998} Workshop, 
(Obs. Paris/IAP, Paris, 14 Dec 1998), \\
{\em http://www.iap.fr/users/roukema/CTP98/programme.html} 
\ 
\\ 
(arXiv:astro-ph/0010170)

%
%
%
%


\bibitem{Corn96} 
Cornish N.~J., Spergel D.~N., Starkman G.~D., 1996, 
arXiv:gr-qc/9602039

\bibitem{Corn98a} \epref{Cornish N.~J., Spergel D.~N., 
Starkman G.~D.}{astro-ph/9708225}{1998a}   

\bibitem{Corn98b} \joref{Cornish N.~J., 
Spergel D.~N., Starkman G.~D.}{\cqg}{15}{2657}{1998}
\ (arXiv:astro-ph/9801212) 


%
%
%
\bibitem{DowS98} 
\joref{Dowker H.~F., Surya~S.}{PhysRevD}{58}{124019}{1998} 
\ (arXiv:gr-qc/9711070)

\bibitem{Fag02} Fagundes H.~V., 2002, 
Contribution to the XXII Brazilian National Meeting on Particles and Fields, 
S{\~a}o Lourenco, October 2001





%

\bibitem{Gaus02} 
\joref{Gausmann E., Lehoucq R., Luminet J.-P., Uzan J.-Ph., 
Weeks J.}{\cqg}{18}{5155}{2001}
\ (arXiv:gr-qc/0106033)


\bibitem{LaLu95} \joref{Lachi\`eze--Rey M.,  
Luminet J.--P.}{Phys. Rep.}{254}{136}{1995} 
\ (arXiv:gr-qc/9605010)
%
%
%
%
%
\bibitem{Levin01} 
\joref{Levin J.}{Phys.Reports}{submitted}{arXiv:gr-qc/0108043}{2002} 
 
%
\bibitem{LR99} Luminet J.--P., Roukema B.~F., 1999, 
`Topology of the Universe: Theory and Observations', 
Carg\`ese summer school `Theoretical and Observational Cosmology', ed. 
Lachi\`eze-Rey M., Netherlands: Kluwer, p117 \ (arXiv:astro-ph/9901364)



%
%
\bibitem{Rouk00a} \joref{Roukema B.~F.}{\mnras}{312}{712}{2000a} 
\ (arXiv:astro-ph/9910272)

\bibitem{Rouk00d} \joref{Roukema B.~F.}{Bull.Astr.Soc.India}{28}{483}{2000}
\ (arXiv:astro-ph/0010185)

\bibitem{Rouk01a} Roukema B.~F., 2001, 
in {Proceedings
of the Soci\'et\'e francaise d'astronomie et astrophysique, 2001} 
\ (arXiv:astro-ph/0106272)

\bibitem{RMBS01} Roukema, B.F., Marecki, A., Bajtlik, S., Spencer, R.E., 2001, 
in
{
``Where's the Matter? Tracing Dark and Bright Matter  
with the New Generation of Large Scale Surveys''}, eds.
Treyer \& Tresse, 2001, Frontier Group
\ (arXiv:astro-ph/0111052)


\bibitem{Rouk02} Roukema, B.F. 2002, 
in {Marcel Grossmann IX Conference}, eds Ruffini  et al., p1901 
\ (arXiv:astro-ph/0010189) 


\bibitem{Schw00} 
\joref{Schwarzschild K.}{Vier. d. Astr. Gess.}{35}{337}{1900}~; 
English translation: \joref{Schwarzschild K.}{\cqg}{15}{2539}{1998}

\bibitem{Stark98} \joref{Starkman G.~D.}{\cqg}{15}{2529}{1998} and the rest of
that volume

%
%
%

\bibitem{Wambs02} Wambsganss J., 2002, in
{``Where's the Matter? Tracing Dark and Bright Matter  
with the New Generation of Large Scale Surveys''}, eds.
Treyer \& Tresse, 2001, Frontier Group


\bibitem{Wich00} 
Wichoski U., 2000, in proceedings of the Cosmic Topology in Paris 
1998 workshop, arXiv:astro-ph/0010170

\end{thebibliography}
\end{document}